\documentstyle[prl,aps,floats,epsf,times,latexsym]{revtex}

\def\np#1#2#3   {{ Nucl. Phys.} {\bf#1}, #2 (#3) }
\def\pcps#1#2#3 {{ Proc. Cam. Phil. Soc.} {\bf#1}, #2 (#3). }
\def\pl#1#2#3   {{ Phys. Lett.} {\bf#1}, #2 (#3) }
\def\plc#1#2#3   {{ Phys. Lett.} {\bf#1}, #2 (#3); }
\def\prep#1#2#3 {{ Phys. Rep.} {\bf#1}, #2 (#3). }
\def\prev#1#2#3 {{ Phys. Rev.} {\bf#1}, #2 (#3). }
\def\prl#1#2#3  {{ Phys. Rev. Lett.} {\bf#1}, #2 (#3). }
\def\prs#1#2#3  {{ Proc. Roy. Soc.} {\bf#1}, #2 (#3). }
\def\ptp#1#2#3  {{ Prog. Th. Phys.} {\bf#1}, #2 (#3). }
\def\rmp#1#2#3  {{ Rev. Mod. Phys.} {\bf#1}, #2 (#3). }
\def\rpp#1#2#3  {{ Rep. Prog. Phys.} {\bf#1}, #2 (#3). }
\def\zp#1#2#3   {{ Zeit. Phys.} {\bf#1}, #2 (#3). }
\def\epj#1#2#3   {{ Eur. Phys. Jour.} {\bf#1}, #2 (#3). }
\def\nim#1#2#3   {{ Nucl. Instr. Meth.} {\bf#1}, #2 (#3). }

\newcommand{\rms}{\rm\scriptstyle}
\newcommand{\stw}{\mbox{$\sin^2\theta_W$}}

\newcommand{\nub}{\overline{\nu}}

\newcommand{\nubar}[0]{\overline{\nu}}

\newcommand{\Rnu}{\mbox{$R^{\nu}$}}
\newcommand{\Rnub}{\mbox{$R^{\nub}$}}
\newcommand{\Rmeas}{\mbox{$R_{\rms exp}$}}
\newcommand{\Rmeasnu}{\mbox{$R_{\rms exp}^{\nu}$}}
\newcommand{\Rmeasnub}{\mbox{$R_{\rms exp}^{\nub}$}}

\renewcommand{\Rmeas}{\mbox{$R_{\rms exp}$}}   

\begin{document}

\twocolumn
%

{\bf Reply to the Comment on ``A Precise Determination of Electroweak
Parameters in Neutrino-Nucleon Scattering''}

\vspace{0.05in}
In a recent comment~\cite{thomas-comment}, Miller and Thomas correctly
indicate that if nuclear shadowing effects differed significantly between
neutrino neutral current (NC) and charged current (CC) interactions, this
difference would impact NuTeV's measurement of $\stw$.  
%
%
As motivation, they offer a specific vector meson dominance (VMD)
shadowing model~\cite{VMD} and argue that nuclear shadowing within the
VMD model is weaker for $Z^0$ exchange than for $W^{\pm}$
exchange~\cite{VMDsize}, thereby increasing the predictions for $\Rnu$
and $\Rnub$ for a portion of the NuTeV data in the low $Q^2$ shadowing
region. This effect of VMD models, though, has the {\em wrong sign}, since 
NuTeV measures ratios for neutrino and antineutrino scattering processes,
$\Rmeasnu$ and $\Rmeasnub$, that are both smaller than expected.
%
%
Furthermore, this class of 
models in general and the specific model offered, both fall well short of 
providing an explanation for the NuTeV observations for a number of reasons.

First, shadowing by VMD is not supported by charged-lepton 
deep inelastic scattering (DIS) data in the relevant kinematic regime. 
The VMD model can be tested by looking for deviations from logarithmic 
$Q^2$ dependence of shadowing, particularly at low $Q^2$, where VMD models 
would predict a $Q^2$ dependence of the form $1/(Q^2+m_V^2)$, $m_V$ 
being the mass of the vector meson.  The most precise data which overlaps 
NuTeV's kinematic region
%
%
($97\%$ of the NuTeV data is contained within $1<Q^2<140$~GeV$^2$, 
$0.01<x<0.75$) comes from the NMC experiment, which observes only the
logarithmic $Q^2$ dependence predicted by perturbative QCD
~\cite{nmc}. 
%
%
%
%
The lack of evidence for strong $Q^2$ dependence of shadowing suggests
that the conventional modeling of shadowing as a change in parton
distribution functions is appropriate in the NuTeV kinematic
region~\cite{pomeron}.
Significant $Q^2$ dependence is observed in charged-lepton scattering  
at low $Q^2$, but in a region irrelevant for NuTeV.

Miller and Thomas do not supply a theoretical framework 
for evaluating the impact of the model on the NuTeV results, nor are their 
estimates
relevant for the NuTeV  kinematics (the mean NuTeV $Q^2$ is 
$25.6$~GeV$^2$ for $\nu$ events and $15.4$~GeV$^2$ for $\nubar$ events).
%
%
%
%
We attempt to apply their model by including the small effect of 
VMD neutral current shadowing and 
using the $x$ dependence 
as given at $5$~GeV$^2$ in Ref.~\cite{Boros}, 
scaled with the above $Q^2$ dependence assuming $m_V=m_\rho$ 
(the 
lowest possible vector meson mass gives the maximum effect). We 
find that the predictions for $\Rmeasnu$ and $\Rmeasnub$ increase 
by $0.6\%$ and $1.2\%$, respectively. 
%
%
While these numbers are
somewhat consistent with the $\epsilon$ and 
$\overline{\epsilon}$ values~\cite{epsilon} which Miller and Thomas claim 
account for the entire NuTeV discrepancy, they neglect to include the high 
degree of correlation between the neutrino and antineutrino ratios in 
evaluating the effect on $R^-$. This brings us to the final point.


%
%
Differences in neutral and charged current shadowing will affect
$\Rnu$ and $\Rnub$ significantly.  However, the NuTeV result derives
$\sin^2\theta_W$ from the Paschos-Wolfenstein
$R^-=(\Rnu-r\Rnub)/(1-r)$, where $r$ is the ratio of $\nubar$ to $\nu$
charged current cross-sections ($r\approx1/2$). 
This approach was chosen by NuTeV to limit sensitivity
to suppression of charged current production of charm quarks from
scattering on the strange sea, yet it is also equally effective at
reducing sensitivity to other parts of the cross-section common
to $\nu$ and $\nubar$, such
as $R_L$ or differences in neutral and charged current nuclear
shadowing.  A low $x$ phenomenon like nuclear shadowing affects
primarily sea quark cross-sections which contribute equally to
$\nu$ and $\nub$ cross-sections.  For this reason, the $\epsilon$ and
$\overline{\epsilon}$ parameters introduced in the
comment~\cite{thomas-comment} will be highly correlated, regardless of
the specifics of the shadowing model, and will affect $R^-$ only slightly. 
The NuTeV
data themselves therefore rule out {\em any} such explanation because of the
enormous shifts in $\Rnu$ and $\Rnub$ individually required to induce
a significant shift in $R^-$.


Even if we arbitrarily {\em increase} the effect of the VMD model in order
to explain the NuTeV $R^-$, NuTeV's
separate measurements 
of $\Rmeasnu$ and $\Rmeasnub$ still cannot be accommodated.  
Defining $\Delta R=\Rmeas-\Rmeas^{\rms (SM)}$, NuTeV has measured:
\begin{eqnarray*}
\Delta R^{\nu}=-0.0032\pm0.0013=\Rmeasnu\times(0.9919\pm0.0033),\\
\Delta R^{\nubar}=-0.0016\pm0.0028=\Rmeasnub\times(0.9960\pm0.0069),
\end{eqnarray*}
with a correlation coefficient between the uncertainties of $0.638$.
This VMD model would therefore increase the discrepancy in $\Rmeasnu$
from $2.5$ to $4.5$ standard deviations and in $\Rmeasnub$ from $0.6$
to $2.3$ standard deviations, and clearly a larger effect would be
more disfavored.  Also, note that the VMD model predicts
the largest change in $\Rmeasnub$ whereas NuTeV
observes a discrepancy primarily in $\Rmeasnu$. 
%
%

In conclusion, the VMD mechanism used in this
comment~\cite{thomas-comment} to motivate the possibility of a large
difference in neutral and charged current shadowing is not motivated
by charged lepton DIS data in the NuTeV kinematic region, and would have a
smaller effect on the NuTeV measurement than is stated.  Furthermore,
VMD shadowing of sufficient size to explain the NuTeV $\stw$ is 
conclusively excluded due to inconsistency with the NuTeV data itself.
More
generally, because any model of different neutral and charged current
nuclear shadowing will change $\Rmeasnu$ and $\Rmeasnub$
more than $R^-$, it is unlikely that any such model could explain the
discrepancy in NuTeV's measurement of $\stw$. 

%


\begin{references}
\bibitem{VMDsize} In VMD models, the effect on $Z^0$ exchange, as a fraction 
                  of the cross-section, is about $1/3$ of that in $W^\pm$ 
                  exchange, which is in turn about $1/2$ of that in $\gamma$ 
                  exchange.
\end{references}
\end{document}